\documentclass[11pt]{article} 
\usepackage{changebar}
\usepackage{graphicx}
\usepackage{epsfig}
\usepackage{subfigure}
\usepackage{amsmath}
\usepackage{rotating}
\usepackage{amsfonts}
\usepackage{amssymb} 
\begin{document}
\title{Linear    response subordination to  intermittent energy release in off-equilibrium aging dynamics}
\author{Simon Christiansen and Paolo Sibani\\
Institut for Fysik og Kemi,  SDU, DK5230 Odense M}

\date{\today}
\maketitle
\begin{abstract} The interpretation of  experimental and numerical 
 data describing  off-equilibrium aging dynamics crucially depends on the 
connection between  spontaneous and  induced fluctuations.  The hypothesis 
that linear response fluctuations  are statistically subordinated to  irreversible outbursts 
of energy, so-called \emph{quakes}, leads to predictions for averages
and fluctuations spectra of physical observables 
in reasonable agreement with experimental results [see e.g. Sibani et al.,  Phys. Rev. B74:224407, 2006]. 
Using simulational data from   a  simple but 
representative Ising  model with plaquette interactions,
direct statistical evidence supporting the hypothesis is presented and discussed in this work.
 A  strict  temporal correlation between  quakes
and   intermittent  magnetization fluctuations is demonstrated. The
external magnetic  field is shown to   bias   the pre-existent intermittent tails of 
the magnetic fluctuation distribution, with little or
no  effect on the Gaussian part of the latter. Its  impact  on  
energy fluctuations is  shown to be negligible.  
 Linear response is thus controlled by the quakes and  
inherits  their temporal statistics.  
These   findings  provide a theoretical basis for analyzing 
intermittent linear response data from aging system  in the same way as  thermal energy 
fluctuations, which are far more difficult to measure. \\
{\bf pacs} 61.43.Fs,65.60.+a,05.40.-a
 \end{abstract} 

\section{Introduction}  
How spontaneous fluctuations 
and  linear response are related in off-equilibrium thermal dynamics
is an  open problem of considerable interest, as  linear response measurements~\cite{Svedlindh87,Vincent96,Jonason98,Komori00b}
are the main in-road into the  rich  phenomenology of aging systems. Highlighting this  issue  
are recent  observations that aging dynamics is intermittent~\cite{Bissig03,Buisson03,Sibani05,Sibani06a}:
  rare but large  
fluctuations with an exponential size distribution  punctuate much smaller 
equilibrium-like  fluctuations with a Gaussian size distribution.   
Simulational~\cite{Oliveira05,Sibani06,Sibani06b,Sibani07}  and experimental~\cite{Sibani06a} evidence in
different areas  supports the idea that  intermittent  changes of   magnetization and other observables  
   are  induced by, and hence statistically
subordinated to,   intermittent and irreversible outburst of heat, so 
called \emph{quakes}. 
 The same hypothesis   leads  to  a  
 widely discussed  asymptotic logarithmic  time re-parameterization of the aging dynamics 
(see e.g. Ref.~\cite{Castillo03}). To ascertain whether or not other mechanisms than subordination  
could produce these  effects, direct  statistical evidence is  called for.  
 
In the following, statistical subordination  and closely related   issues are  numerically 
investigated,  using as a test-bed   an Ising model,    
which is simple and yet constitutes   a \emph{bona fide} instance of a complex aging 
system~\cite{Sibani06b,Lipowski00,Swift00}.    
Intermittent changes 
 in   magnetization  are shown to obey the same  statistics as  the  quakes, and  the magnetic field 
is shown to have negligible influence on the energy relaxation.  Together, these two findings   
imply   that    aging dynamics is quake-driven.    
The   fluctuation statistics is then investigated in  detail, with   
 the model properties in good   agreement with  
previous investigations of intermittent heat flow~\cite{Sibani05,Sibani06b} and of 
magnetic linear response intermittency~\cite{Sibani06a,Sibani07}.   
Quakes in this model are shown to be nearly uncorrelated events, which are   described by a Poisson 
process  whose average  increases    logarithmically in time. Additionally,  
a linear  system-size  dependence of the    
average number of quakes   is  found. This confirms 
that quakes are spatially localized events, a property 
  also   found directly  in this work via  a real space analysis. The  temperature
  dependence of  the   average number of quakes  is 
very weak, except at the lowest temperatures. This hints to  a  hierarchical 
structure in the energy landscape of the thermalized domains spawning the quakes.  
\begin{figure}[t]
$ 
\begin{array}{lr}
\includegraphics[width=.45\textwidth]{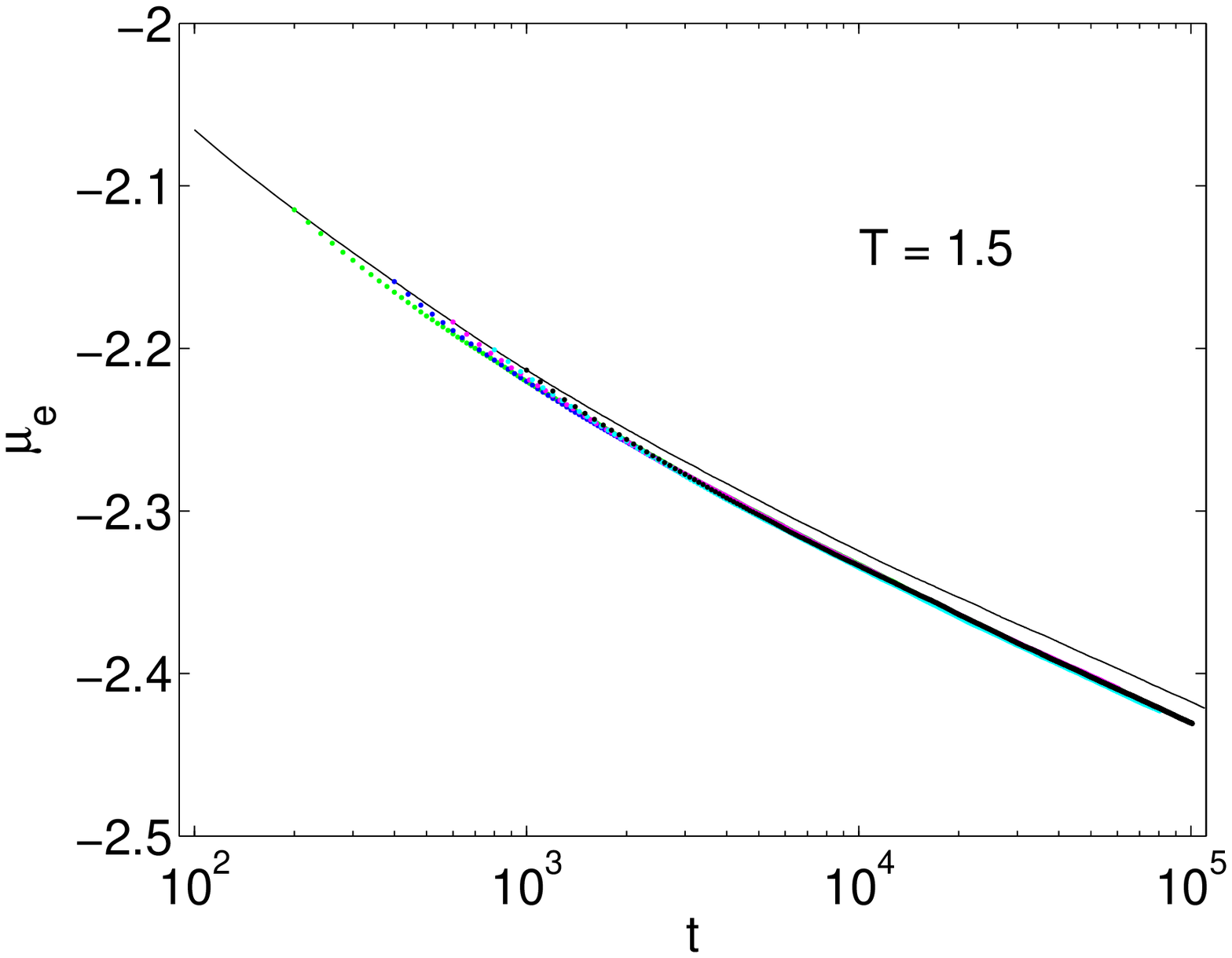} &  
\includegraphics[width=.45\textwidth]{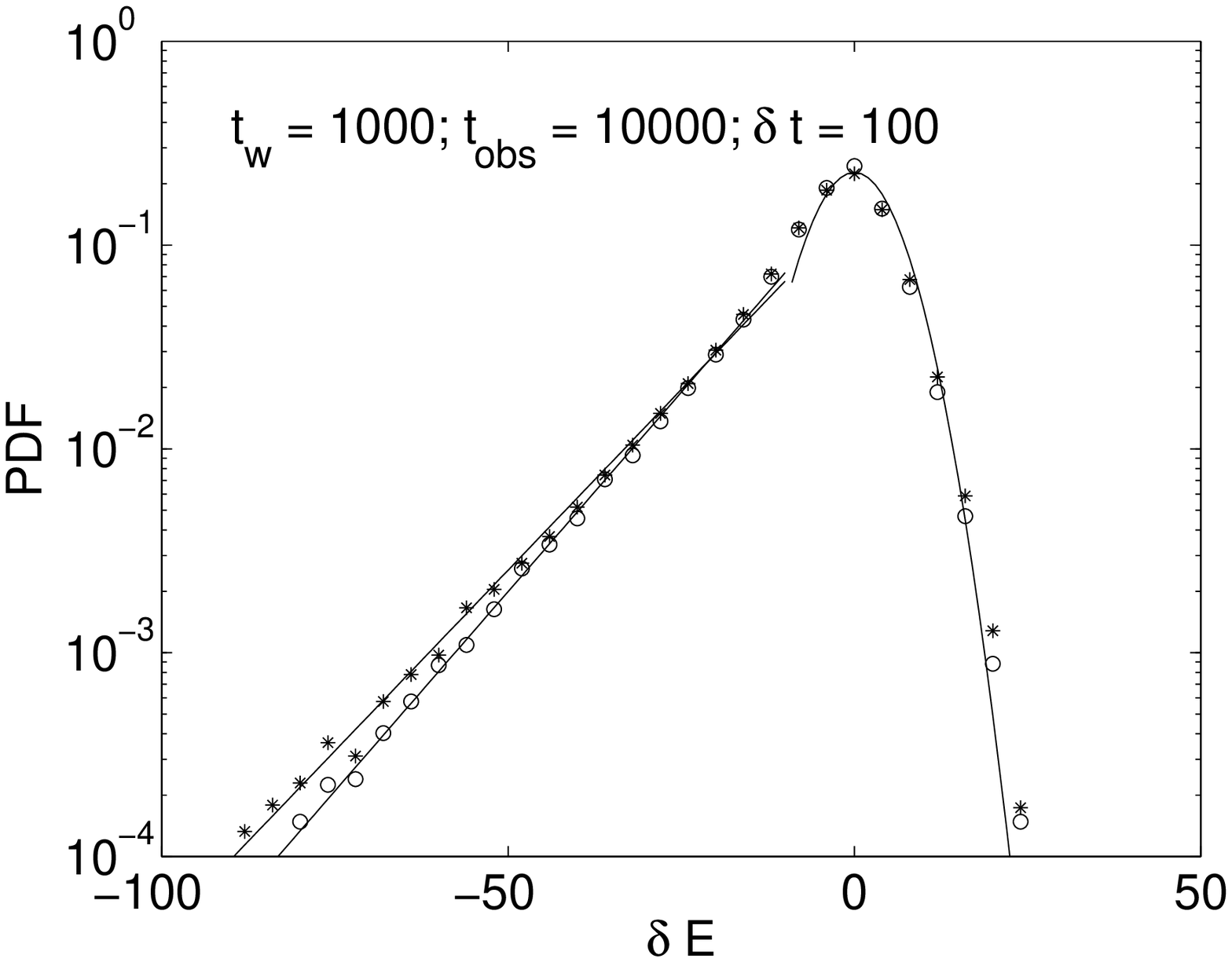}
\end{array}
$  
\caption{
{\em (left)}: (colour on line) The  average energy per spin, $\mu_e$,  is plotted versus  the 
system age $t$,  
with the magnetic field switched on at   times $t_w=200,400,600,800,1000$ and $t_w=2000$.  
The magnetic field has a quite small effect on $\mu_e$. 
{\em (right)}: The  PDF of the energy changes  $\delta E$   
  over a time $\delta t=100$, with (stars) and without (circles) 
  a magnetic field. Both PDFs feature   a central  Gaussian  part of zero 
  average  and  intermittent tails. The  PDFs are nearly identical, except for the  largest and  
  rarest events.  The simulation temperature is $T=1.5$ 
  for all  plots.
 } 
\label{heat_flow}
\end{figure} 
\section{Model and method}
\label{model}  
In the  model, $N$   Ising spins, $\sigma_i = \pm 1$ are  
placed  on a cubic lattice with periodic boundary conditions. They   interact through  
the plaquette Hamiltonian  
\begin{equation}
{\cal H} = -\sum_{{\cal P}_{ijkl}} \sigma_i\sigma_j \sigma_k \sigma_l +  H \eta(t_w -t) \sum_i \sigma_i.
\label{hamilt}
\end{equation}
\begin{figure}[t]
$ 
\begin{array}{lr}
\includegraphics[width=.45\textwidth]{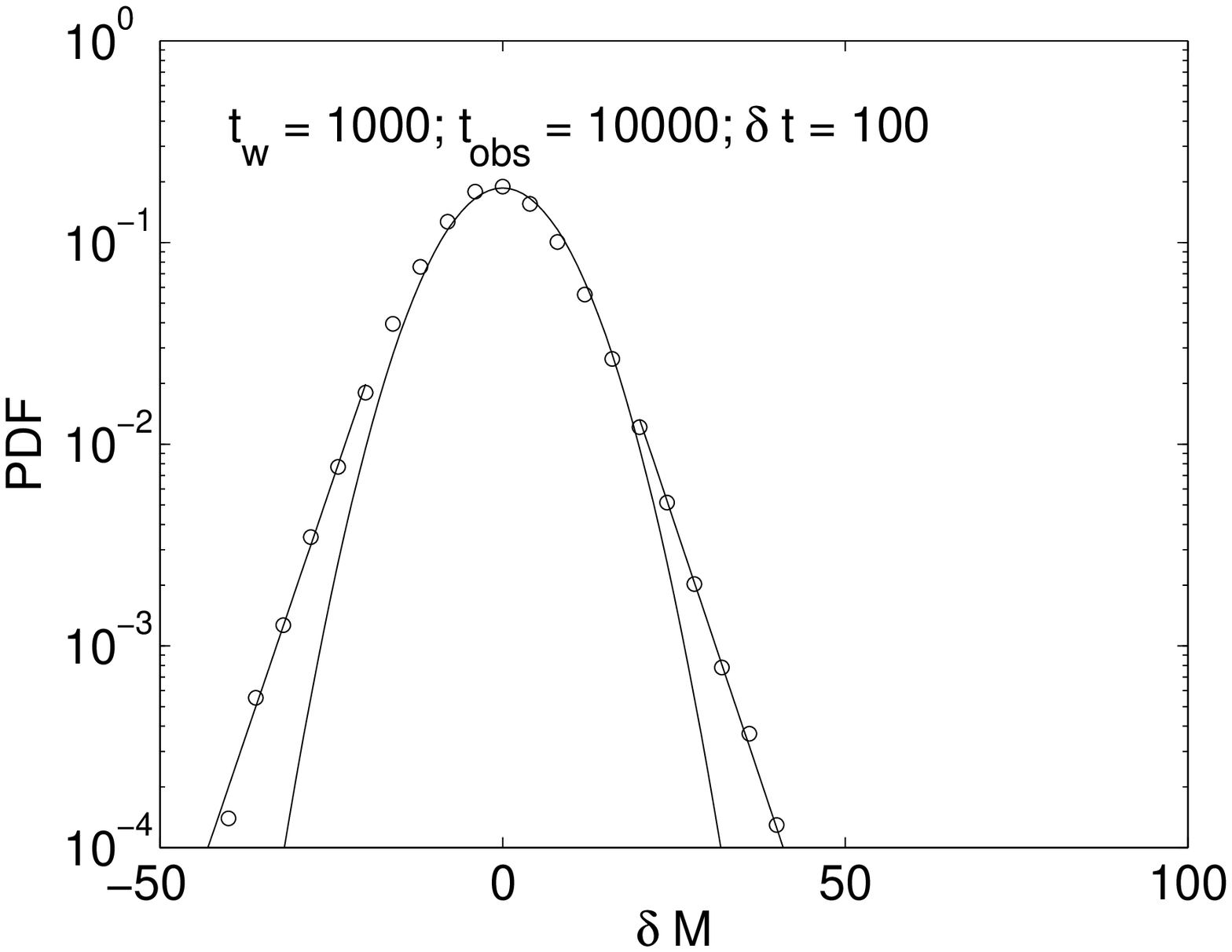} &  
\includegraphics[width=.45\textwidth]{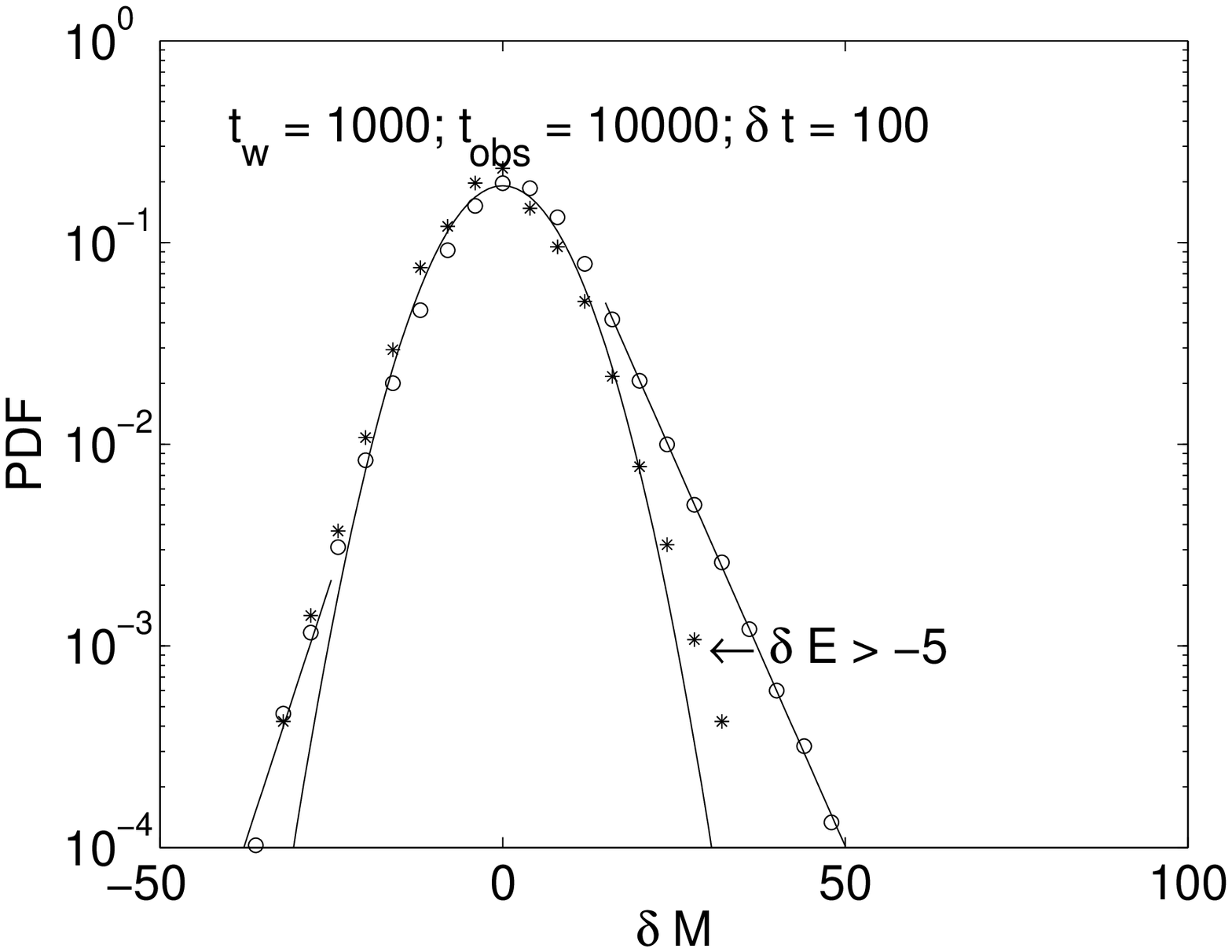}
\end{array}
$  
\caption{(colour online) {\em (left)}:  The PDF of the spontaneous magnetic fluctuations
 has a Gaussian central part and two symmetric intermittent wings. Data are sampled
 in the interval $t_w, t_w + t_{\rm obs}$.
{\em (right)}: The same quantity  (outer graph, circles) when  a field $H=0.3$ turned on 
at $t_w = 1000$. The left intermittent wing is clearly reduced relative to the no-field case, and the right intermittent
wing is correspondingly amplified. The inner, almost Gaussian shaped, PDF (stars)  is the \emph{conditional} PDF obtained by
excluding   the magnetic fluctuations which happen in unison with the quakes. 
 }
\label{mag_PDF}
\end{figure}  
The first sum  runs over the elementary  plaquettes 
of the lattice,  including   for each the  product of the  four  spins 
  located at its corners. 
The second term describes  the  coupling of the total magnetization  $\sum \sigma_i $ to  an external magnetic field. As expressed by the Heaviside step function $\eta(t_w -t)$,
the field   changes instantaneously at  $t=t_w$ from zero to $H>0$.
Previous investigations of the model's properties  in zero field  show
 a low temperature aging regime~\cite{Lipowski00,Swift00},
 during which   energy leaves the system 
 intermittently,  and  at a rate falling off as the inverse time~\cite{Sibani06b}. 

The  present   simulations are  all performed within the aging regime,
 i.e. in the  temperature  range $0.5<T<2.5$, using the rejectionless
  Waiting Time Algorithm (WTM)~\cite{Dall01}.   The `intrinsic'   time unit
  of the  WTM  approximately corresponds to one Monte Carlo sweep. 
 By choosing a   high energy random configuration as initial state for low temperature isothermal
 simulations, an   effectively instantaneous  thermal quench is  achieved.
\begin{figure}[t]
$ 
\begin{array}{lr}
\includegraphics[width=.45\textwidth]{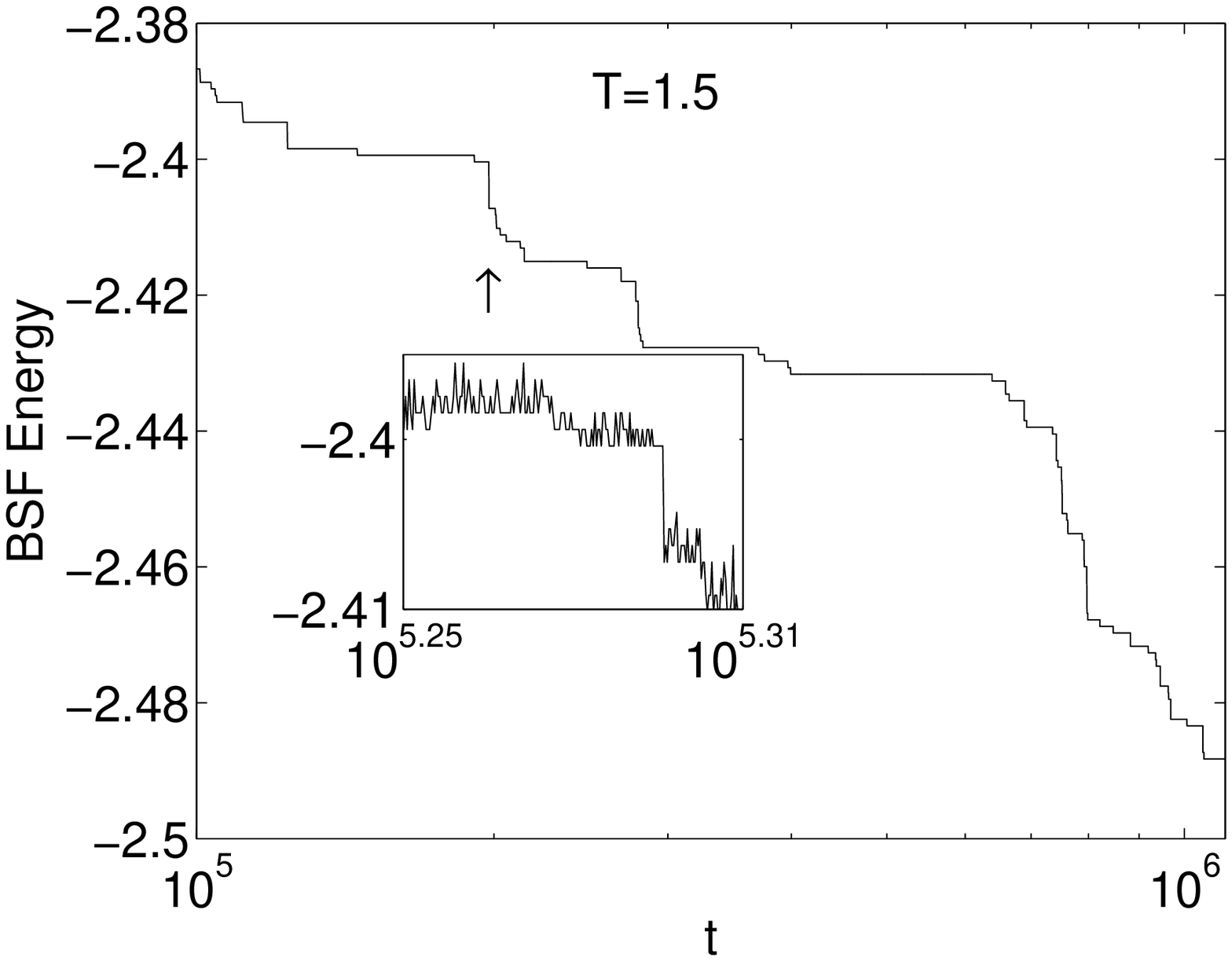} &  
\includegraphics[width=.45\textwidth]{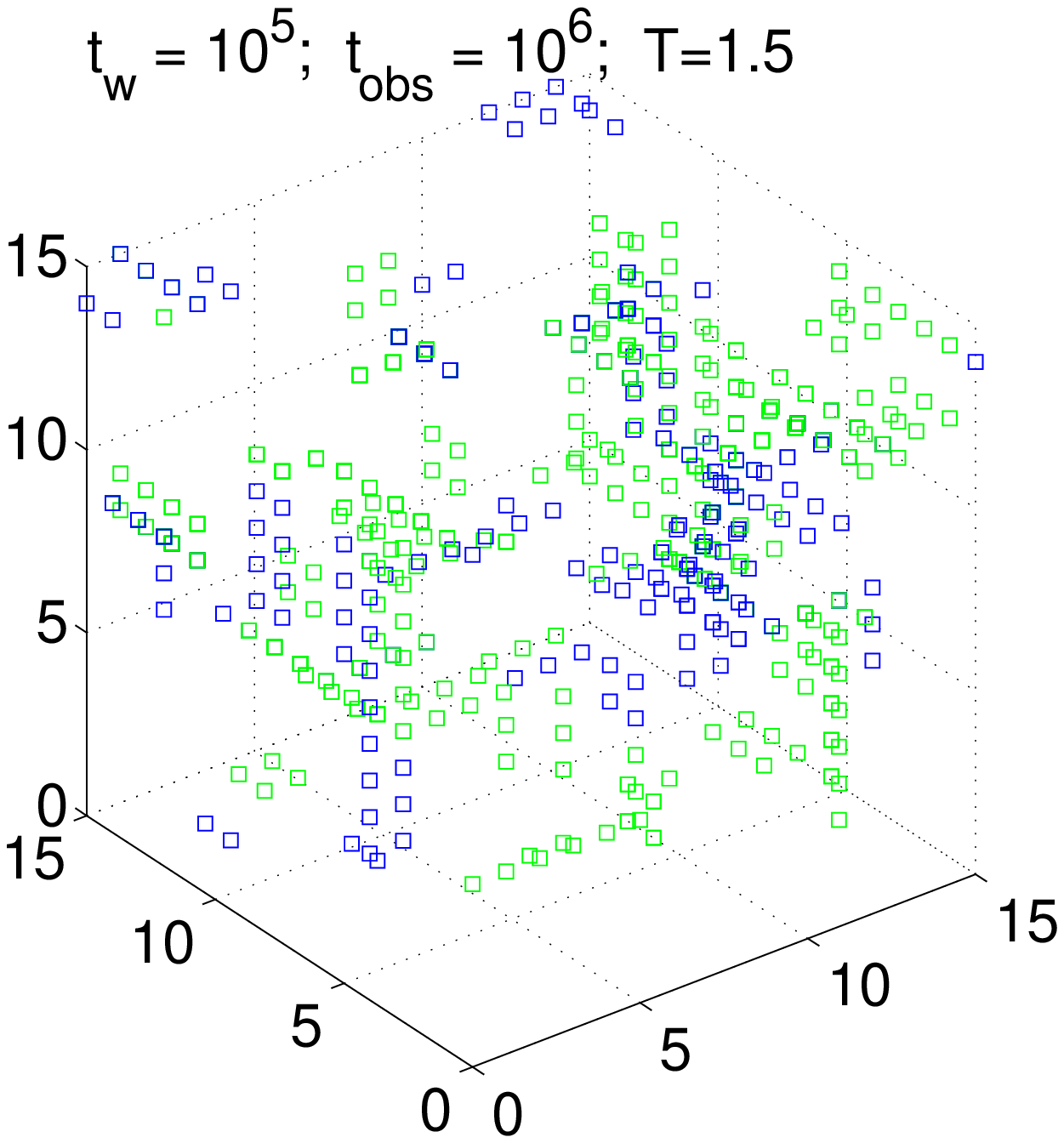}
\end{array}
$  
\caption{(colour online) {\em (a)}:  The record signal, or Best So Far energy, for a simulation 
at $T=1.5$ and no field. The full signal, which  includes reversible fluctuations, is shown in the insert
for a small sub-interval of the observation time.  
{\em (b)}:   active spins participating in any  quakes occurring during the
interval $[10^5,1.1 \times 10^6]$ are shown in colour. All other spins are omitted.
By definition,  active  spins   change their absolute 
 average value  over five consecutive time units 
by  exactly $2$ (blue) or by   $1.6$ (green). 
Most spins are excluded as they either  fluctuate with no  average change or 
retain a fixed value during all $\delta t$'s.
 }
\label{aging_of_energy_and_quakes}
\end{figure}    
  For each set of physical
parameters, Probability Distribution Functions (PDFs)  are collected over  $2000$ 
independent runs, and other statistical data  over $1000$ independent runs.
 The symbol $t$ stands for the time elapsed from the initial quench (and from the beginning of the
 simulations). The symbol $t_w$ is  the time at which the field is switched on,
 while  $t_{\rm obs} = t - t_w$ stands 
 for the `observation' time, during which data are collected.
 \begin{figure}[t]
$ 
\begin{array}{lr}
\includegraphics[width=.45\textwidth]{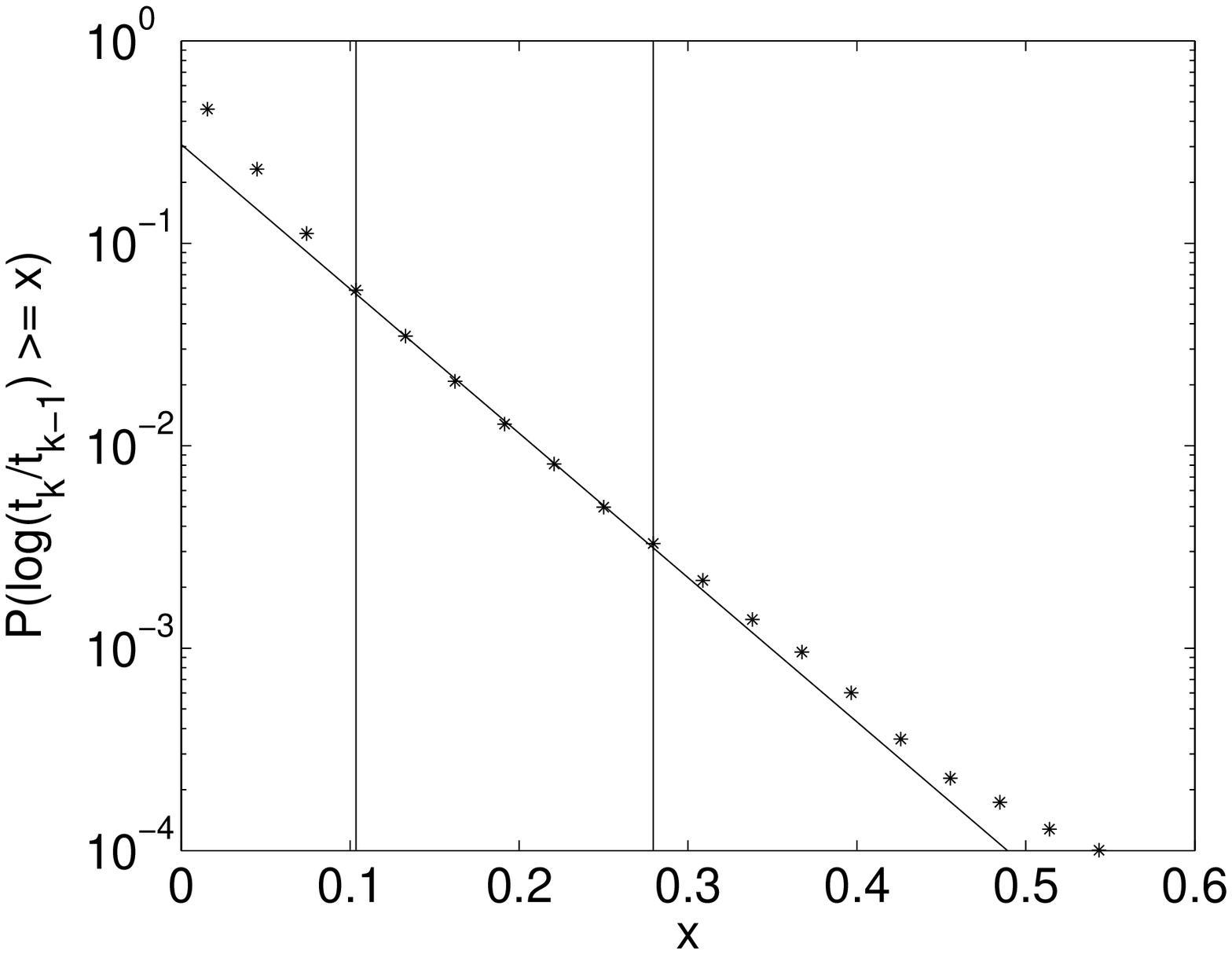} &  
\includegraphics[width=.45\textwidth]{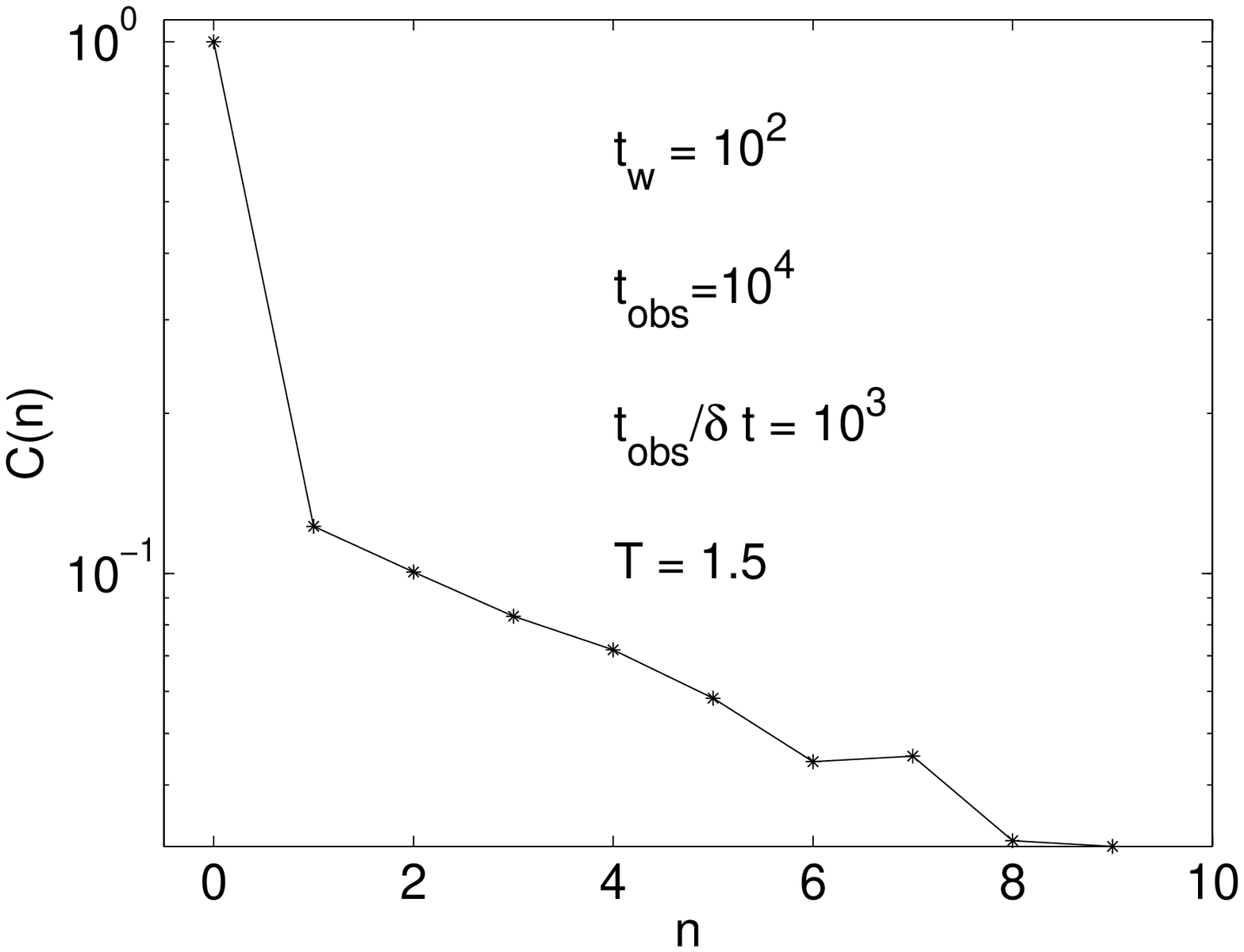}
\end{array}
$ 
\caption{(colour online) {\em (a)}:  The empirical  cumulative distribution (stars) of the 'logarithmic waiting times'
$\tau_k = \ln(t_k) - \ln(t_{k-1})$ is plotted on a log scale. The full line is a least square fit to $y \propto \exp(-\alpha x)$,
using the data points between the vertical lines.
{\em (b)}: The normalized  correlation between 
  $\tau_k$ and $\tau_{k+n} $ (stars) is plotted versus $n$ on a logarithmic scale.
  The line is a guide to the eye.  We see that consecutive data points
  ($n=1$) are only weakly correlated, and that the correlation decays exponentially for $n>1$. This is in reasonable 
  qualitative agreement with the theoretical approximation $C(n) = \delta_{n,0}$.   
 }
\label{logwaitingtimes}
\end{figure}
The value of the external field is set to  $H=0.3$ for $t>t_w$. 
  The thermal energy is denoted by $E$, and the magnetization 
 by $M$. The average energy per spin is denoted by $\mu_e$. 
 The PDF of fluctuations in energy and magnetization
 are constructed using  finite time differences of $E$ and $M$,  taken  over
 short time  intervals of length  $\delta t \ll t_{\rm obs}$.
\section{Results}  To qualify as a  probe, an applied perturbation must not   significantly change 
the  dynamics of the   system investigated.  Specifically for the present  model,    the 
 average energy vs.  time 
should be nearly unaffected. Secondly, fluctuation  spectra    should not  undergo qualitative changes. 
Figure~\ref{heat_flow}, left panel,  shows  the average energy  as a function of time, 
with six data sets  corresponding to   the perturbation switched on
at $t_w$  values ranging from $t_w=200$ to $t_w=2000$.
As expected, the  field has little impact on the energy.  The right panel  shows, on a logarithmic scale, 
the PDF of the energy fluctuations in zero field (circles) and 
when a field is switched on   at $t_w = 1000$ (stars).   Gaussian energy fluctuations of zero average
are flanked on the left by an intermittent tail, which carries  the net heat flow out of the
system. Again, only a  minor  difference is seen between the unperturbed and perturbed
fluctuation spectra,  and only   for  the largest and rarest of the fluctuations.  
Since  the magnetic contribution   to
the average energy is  negligible, quakes    the dissipation of  the excess energy entrapped in the 
initial configuration, see Fig.~\ref{heat_flow}, in the same   as 
in an  unperturbed system would do~\cite{Sibani06b}.
 
While a  magnetic field  induces a   non-zero average magnetization, 
 it does not change the overall  structure  of the  
 spectra:  The left panel of Fig.\ref{mag_PDF} shows the PDF of the spontaneous magnetic fluctuations 
 occurring in the interval 
$[t_w,t_w+t_{\rm obs}]$. Intermittent wings  symmetrically  
extend the central Gaussian part of the PDF. The \emph{outer} curve (circles) in the right panel 
of the same figure   depicts  the PDF obtained  when the  field is turned on at $t_w=1000$.
The  positive  intermittent tail  is  enhanced,  the negative tail is 
reduced  and the Gaussian part is not affected.
Thus, the   net average magnetization induced by the field, 
i.e. the linear response,  arises through a biasing effect on the distribution of the spontaneous  intermittent magnetic fluctuations. 

A key aging  feature~\cite{Sibani05,Sibani06b} is  that the excess energy trapped
by the initial quench leaves  the system through  intermittent quakes. 
Intermittent   magnetic fluctuations  
  have a close temporal association to the quakes:  The 
 inner curve (stars) of Fig.~\ref{mag_PDF}  depicts  a   \emph{conditional} PDF obtained by filtering 
 out the  magnetic fluctuations   which occur   simultaneously
(i.e., in practice,  either within the same    or within  the immediately following $\delta t$ )
with  energy fluctuations of magnitude  $\delta E \le -5$.  The  filtering threshold   
utilized  is near the onset of the  intermittent behavior of the heat flow PDF, 
as seen in Fig~\ref{heat_flow}. 
The filtering produces a  nearly Gaussian PDF, demonstrating that 
quakes and intermittent magnetic fluctuations are  synchronous events.  
In summary, the magnetic fluctuations which 
 contribute to the linear response are subordinated  to the quakes which dissipate the 
excess energy stored in the initial configuration. 
 
 Visual inspection shows that energy traces of the p-spin model have  a fluctuating part 
 superimposed onto   a  monotonic step-wise decay, the latter  given by the function 
 $r_E(t) = \min_{y<t} E(t)$. This  function is  called 
record signal, or Best So Far (BSF) energy.       
 The BSF energy is plotted in the main left panel of Fig.~\ref{aging_of_energy_and_quakes}.
The insert shows the full trace for a shorter interval of time. 
The right panel of Fig.~\ref{aging_of_energy_and_quakes} is a pictorial
rendering of the real space positions  of  spins  
 assessed to participate in one  or more quakes  during the  
 time interval $[10^5,1.1\times 10^6]$. In the plot, the `active'  spins  are shown in colour, and all  others  are omitted.
By definition,     spins  are labeled  as active if the time average over five consecutive time units
of their orientation has changed  
by exactly    $2$ units (blue) or by $1.6$ units (green).  
   Note that the  system-spanning   
 spatial distribution of the involved spins   integrates  the effect of the many unrelated quakes occurring in  
the observation time. Any single  quake  typically involves small clusters only.

The quake statistics of this model should be  amenable to a description as  a Poisson  
process  with   average~\cite{Sibani93a,Sibani03,Anderson04,Sibani05} 
\begin{equation}
n_I(t',t)  = \alpha \ln (t/t'). 
\label{average}
\end{equation}   
Assuming that quakes can be treated as instantaneous events, 
which occur at times $t_1, \ldots t_q \ldots$,  
a testable property equivalent to  Eq.~\ref{average},  
 is that the logarithmic differences  
(rather than  linear  differences) $\tau_q \stackrel{\rm def}{=} \ln(t_q/t_{q-1})$
 are  independent and identically distributed
stochastic variables, sharing   the exponential distribution
\begin{equation}
{\rm Prob}(\tau_q > x)   = \exp(-\alpha x). 
\label{distrib}
\end{equation}  
 A simple choice  to evaluate the $\tau_q$ is to identify the quake
 times $t_q$  with the  steps 
of   the BSF energy. Admittedly, this introduces  an 
unwanted  dependence  on   the data  sampling frequency, when  the  latter 
 is too high:  the same   transition 
 from one metastable  configuration to another may then  be registered 
as  multiple  quakes. This leads to over-counting,   and 
as discussed below,   to spurious  correlations  appearing  between  successive quakes. 
 \begin{figure}[t]
$ 
\begin{array}{lr}
\includegraphics[width=.45\textwidth]{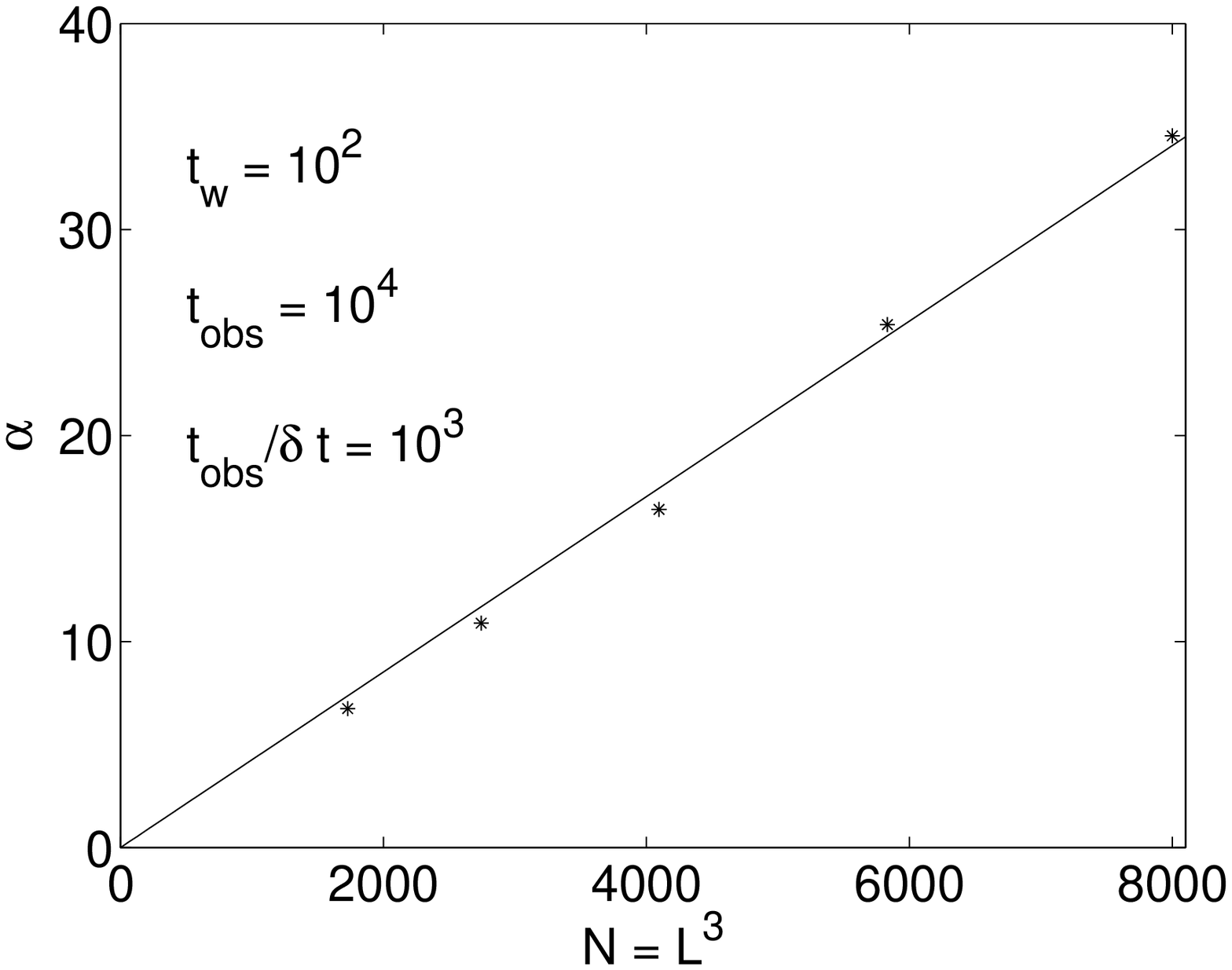} &  
\includegraphics[width=.45\textwidth]{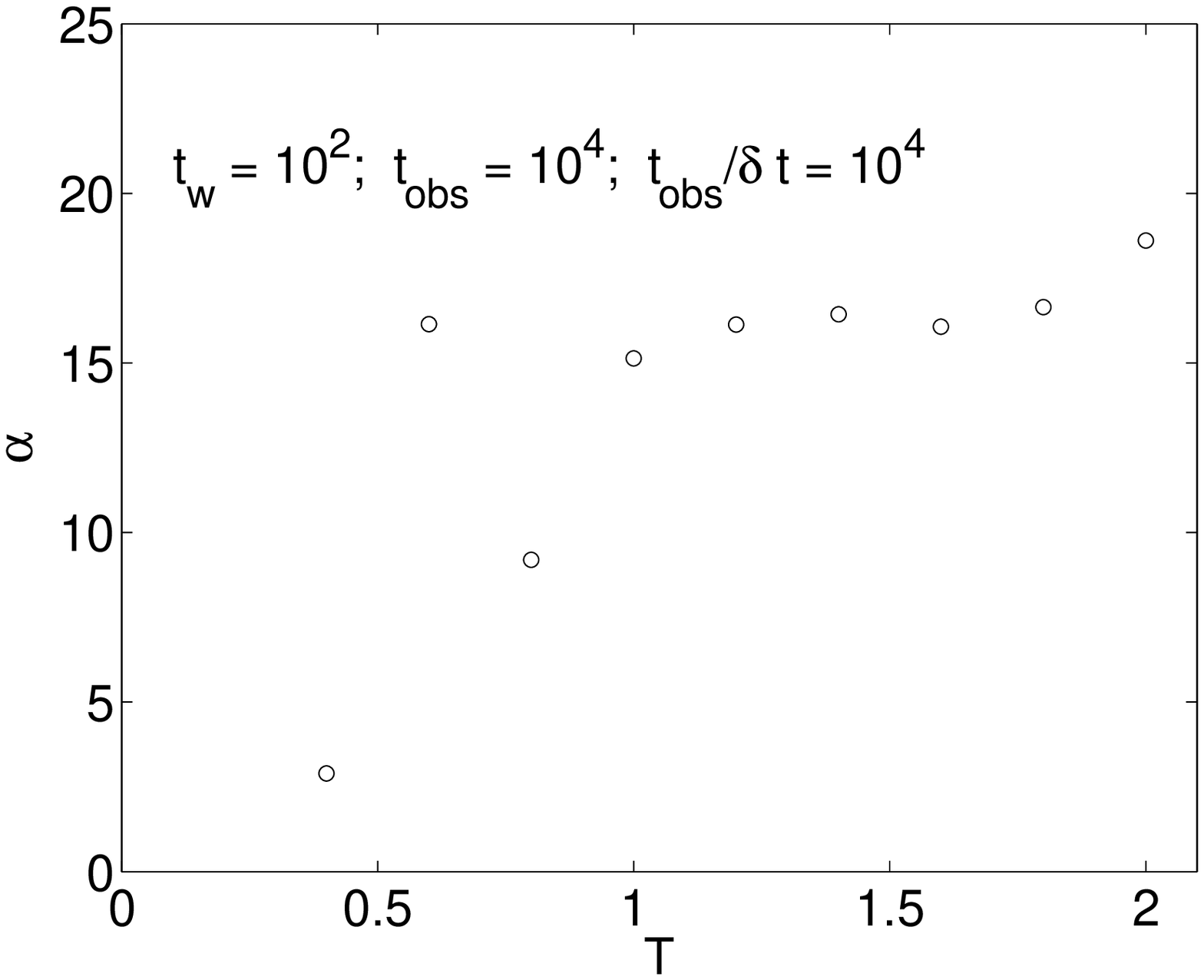}
\end{array}
$  
\caption{(colour online) {\em (a)}:  The logarithmic rate of quakes $\alpha$  is plotted versus   system size
for $T=1.5$.
{\em (b)}: The  same quantity is plotted versus $T$,  for $N=16^3$.
The values of $\alpha$ are estimated from an exponential fit to the distribution of logarithmic waiting times, 
 see Fig.~\ref{logwaitingtimes}a.
 }
\label{alpha_vs_TandN}
\end{figure}   

From the   empirical series of $\{\tau_q\}_{q=1,2,\ldots}$ collected  for 
each trajectory,   the  correlation function is estimated  as  
$C(k)=\langle \tau_q \tau_{q+k}\rangle_q - \langle \tau_q \rangle_q \langle \tau_{q+k}\rangle_q$.  
The result is  then averaged over all 
trajectories.   Systematic and statistical errors
 in the quake  identification procedure  lead to 
 deviations    from the theoretically expected  
Kronecker delta   $C(k)/C(0)= \delta_{k,0}$. 

In the  left panel of Fig.~\ref{logwaitingtimes}  the  distribution of the $\tau_q$  
(dots) is   well approximated by an exponential decay over nearly  three decades (line). 
 For  large values of the abscissa,  deviations  likely stem from a statistical  under-sampling of rare
 events. For  $x \approx  0$ (the first three points are excluded from the fit) deviations 
 indicate  that successive intervals of equal  or near equal duration   
appear   more frequently    than theoretically expected. This can arise 
when the interval between successive events is spuriously affected by
the sampling time.  The correlation function $C(k)$ of the logarithmic time differences,
normalized to $C(k=0)=1$,   is shown in   the right panel of the figure.  
After   a  drop    to approximately $1/10$ at $k=1$, the function  tapers off 
exponentially with increasing $k$. The  (modest)  residual correlation confirms  
that   successive events are occasionally   mis-classified as quakes. 
 
A spatially extended glassy system with short range interactions
is expected to contain 
a  number, say  $\alpha$,   of independently relaxing thermalized domains, 
with  a small  and slowly growing characteristic size~\cite{Rieger93,Komori99}.
Assuming that    quakes  originate independently within the domains,  
 the observed fluctuations  integrate the effect of  $\alpha$ independent Poisson processes,
whence $\alpha$  equals the  logarithmic rate of quakes.    
Equivalently, its value can    be   estimated from the  
exponential  distribution of the logarithmic waiting times $\tau_q$, see Fig.~\ref{logwaitingtimes},
as done  in the present case.

The number of domains,   and hence
 $\alpha$,  must increase linearly with the system size, as 
 demonstrated in the left panel of Figure \ref{alpha_vs_TandN}.
The temperature dependence (or lack thereof) of $\alpha$ is related to 
the  geometrical properties of the configuration space, or energy landscape, of each domain.  
 For record sized thermal fluctuations to  elicit
attractor changes,  the energy landscape must be 
scale invariant~\cite{Sibani07}. As a consequence, 
changing the temperature should  not change  
 $\alpha$ at all. Note however that  
scale invariance  cannot hold  below  a  cut-off value 
where the   granularity  of the energy values makes itself felt.
In  the present model, the numerically smallest  energy change following 
a single spin flip is $\delta E = \pm 4$, i.e. unlike models
with Gaussian quenched disorder, the granularity is important. 
The right  panel of Figure \ref{alpha_vs_TandN} shows that $\alpha$ 
has a  modest temperature dependence for $1\le T\le 2$, i.e. for a major part of the range where
aging behavior is observed. For lower  temperatures, a clear  $T$ dependence is visible.  

\section{Summary and conclusions} 
Direct numerical evidence has been provided that 
 intermittent magnetic fluctuations  are statistically  subordinated to a 
certain type of events,  quakes, which   dissipate   the excess energy trapped in the 
initial configuration.
The   external  field  does not    
  alter   the    temporal statistics of the quakes and of the spontaneous
  magnetic fluctuations. It only   slightly biases the size distribution of the latter.
 Therefore, the field can  rightly 
be  considered as a  probe of the unperturbed off-equilibrium aging dynamics. 
In agreement  with   previous investigations  
of  other models \cite{Sibani07} and  experiments~\cite{Sibani06a},
the temporal statistics of    intermittent energy and magnetization 
changes,    is    shown to be well described 
by  Poisson process. 
 
Considering that aging dynamics is widely insensitive to  details of the
microscopic interactions,  
it seems reasonable to  assume that the above findings are 
valid beyond the plaquette model itself.
It should therefore be  possible to   analyze a wide range of intermittent linear response data from   
complex dynamical systems   precisely as done  for the intermittent   heat flow data in Ref.~\cite{Sibani05,Sibani06b}.

\subsection*{Acknowledgemnts} Financial support  from the Danish Natural Sciences Research Council
is gratefully acknowledged.  The authors are indebted to 
the Danish Center for Super Computing (DCSC)  for computer time on
the Horseshoe Cluster, where  most of the simulations were carried out.

\vspace{1cm} 
\bibliographystyle{unsrt.bst}
\bibliography{SD-meld}

\begin{thebibliography}{10}

\bibitem{Svedlindh87}
P.~Svedlindh, P.~Granberg, P.~Nordblad, L.~Lundgren, and H.S. Chen.
\newblock Relaxation in spin glasses at weak magnetic fields.
\newblock {\em Phys. Rev. B}, 35:268--273, 1987.

\bibitem{Vincent96}
{Eric Vincent, Jacques Hammann, Miguel Ocio, Jean-Philippe Bouchaud, and
  Leticia F. Cugliandolo}.
\newblock Slow dynamics and aging in spin-glasses.
\newblock {\em SPEC-SACLAY-96/048}, 1996.

\bibitem{Jonason98}
K.~Jonason, E.~Vincent, J.~Hammann, J.~P. Bouchaud, and P.~Nordblad.
\newblock {M}emory and {C}haos {E}ffects in {S}pin {G}lasses.
\newblock {\em Phys. Rev. Lett.}, 81:3243--3246, 1998.

\bibitem{Komori00b}
H.~Yoshino T.~Komori and H.~Takayama.
\newblock {Numerical study on aging dynamics in Ising spin-glass models.
  Temperature change protocols }.
\newblock {\em J. Phys. Soc. Japan}, 69:228--237, 2000.

\bibitem{Bissig03}
{H. Bissig, S. Romer, Luca Cipelletti Veronique Trappe and Peter
  Schurtenberger}.
\newblock Intermittent dynamics and hyper-aging in dense colloidal gels.
\newblock {\em PhysChemComm}, 6:21--23, 2003.

\bibitem{Buisson03}
{L.~Buisson, L.~Bellon, and S.~Ciliberto}.
\newblock Intermittency in aging.
\newblock {\em J. Phys. Condens. Matter.}, 15:S1163, 2003.

\bibitem{Sibani05}
P.~Sibani and H.~Jeldtoft Jensen.
\newblock Intermittency, aging and extremal fluctuations.
\newblock {\em Europhys. Lett.}, 69:563--569, 2005.

\bibitem{Sibani06a}
{Paolo Sibani, G.F. Rodriguez and G.G. Kenning}.
\newblock Intermittent quakes and record dynamics in the thermoremanent
  magnetization of a spin-glass.
\newblock {\em Phys. Rev. B}, 74:224407, 2006.

\bibitem{Oliveira05}
{L.P. Oliveira, Henrik Jeldtoft Jensen, Mario Nicodemi and Paolo Sibani}.
\newblock Record dynamics and the observed temperature plateau in the magnetic
  creep rate of type ii superconductors.
\newblock {\em Phys. Rev. B}, 71:104526, 2005.

\bibitem{Sibani06}
Paolo Sibani.
\newblock Mesoscopic fluctuations and intermittency in aging dynamics.
\newblock {\em Europhys. Lett.}, 73:69--75, 2006.

\bibitem{Sibani06b}
{Paolo Sibani}.
\newblock Aging and intermittency in a p-spin model.
\newblock {\em Phys. Rev. E}, 74:031115, 2006.

\bibitem{Sibani07}
{P. Sibani}.
\newblock {Linear response in aging glassy systems, intermittency and the
  Poisson statistics of record fluctuations}.
\newblock {\em Eur. Phys. J. B}, 58:483--491, 2007.

\bibitem{Castillo03}
{ Horacio E. Castillo, Claudio Chamon, Leticia F. Cugliandolo, Jos{\'{e}} Luis
  Iguain and Malcom P. Kenneth}.
\newblock Spatially heterogeneous ages in glassy systems.
\newblock {\em Phys. Rev. B}, 68:13442, 2003.

\bibitem{Lipowski00}
A.~Lipowski and D.~Johnston.
\newblock Cooling-rate effects in a model of glasses.
\newblock {\em Phys. Rev. E}, 61:6375--6382, 2000.

\bibitem{Swift00}
{Michael. R. Swift, Hemant Bokil, Rui D. M. Travasso and Alan J. Bray}.
\newblock Glassy behavior in a ferromagnetic p-spin model.
\newblock {\em Phys. Rev. B}, 62:11494--11498, 2000.

\bibitem{Dall01}
Jesper Dall and Paolo Sibani.
\newblock {Faster} {M}onte {C}arlo simulations at low temperatures. {T}he
  waiting time method.
\newblock {\em Comp. Phys. Comm.}, 141:260--267, 2001.

\bibitem{Sibani93a}
P.~Sibani and Peter~B. Littlewood.
\newblock Slow {Dynamics} from {Noise} {A}daptation.
\newblock {\em Phys. Rev. Lett.}, 71:1482--1485, 1993.

\bibitem{Sibani03}
Paolo Sibani and Jesper Dall.
\newblock {Log-Poisson statistics and pure aging in glassy systems.}
\newblock {\em Europhys. Lett.}, 64:8--14, 2003.

\bibitem{Anderson04}
{Paul Anderson, Henrik Jeldtoft Jensen, L.P. Oliveira and Paolo Sibani}.
\newblock Evolution in complex systems.
\newblock {\em Complexity}, 10:49--56, 2004.

\bibitem{Rieger93}
H.~Rieger.
\newblock Non-equilibrium dynamics and aging in the three dimensional {I}sing
  spin-glass model.
\newblock {\em J. Phys. A}, 26:L615--L621, 1993.

\bibitem{Komori99}
H.~Yoshino T.~Komori and H.~Takayama.
\newblock {Numerical study on aging dynamics in the 3D Ising spin-glass model.
  I. Energy relaxation and domain coarsening}.
\newblock {\em J. Phys. Soc. Japan}, 68:3387--3393, 1999.

\end{thebibliography}
\end{document}